# Modulation of domain wall dynamics in TbFeCo single layer nanowire

**Duc-The Ngo, Kotaro Ikeda, Hiroyuki Awano**

Information Storage Materials Laboratory, Toyota Technological Institute, 2-12-1

Hisakata, Tempaku-ku, Nagoya 468-8511, Japan

## Abstract

We demonstrate the possibility to write and modulate the magnetic domain walls in a TbFeCo single layer nanowire (300 nm width, 150 μm length). To realize this, a tiny magnetic domain was nucleated by an Oersted field produced by a 1.6 MHz pulse current (35 mA in amplitude, 5-40 ns in length) crossed the wire. To write the wall to the wire, a DC current was used to drive the nucleated domain (with two walls in two sides) to the wire in accordance with spin-transfer torque mechanism. A critical current density of $J_c = 3.5 \times 10^{10}$ Am$^{-2}$ was required to control the motion of the walls in the wire. It was found that the size of the domain moving in the wire could be adjusted by either external field or the length of the nucleated pulse current. This could be considered as an important note for writing process in domain wall spin-torque devices, especially, memory elements.

## 1. Introduction

In magnetic data storage devices, the data bit is defined as the direction of the magnetization. The information is stored by switching the magnetization to the desired configuration. In some cases, e.g. hard disk, conventional magnetoresistive random access memory (MRAM), the switching is accomplished by a magnetic field generated by an electrical current, which is referred to Oersted field. The other approach recently concerns with spin transfer effect in which the magnetization can be switched by a spin-polarized electron current. Spin transfer phenomenon involves the possibility to reverse the magnetization in a ferromagnet without the application of an external field. This effect was theoretically predicted in 1996 by L. Berger [1] and J. Slonczewski [2]. It immediately attracted much attention both for its fundamental interest as a new physical effect and its huge potential for applications in novel spintronics [3-6]. When a current flows through a ferromagnet, it is spin-polarized by the magnetization of the ferromagnet and carries an angular momentum. The current remains its polarization and so that the angular momentum can interact with the magnetization in subsequent magnetic layer. The spin current exerts a torque (spin-transfer torque - STT) on the magnetization of the subsequent layer that can lead to precession and reversal. STT-MRAM [7] is a typical application of the spin-transfer torque in magnetic multilayer nanostructure in which the information, which is stored in the magnetization of the film layer, is reversed by using a spin-polarized current.

Motion of the magnetic domain walls (DWs) under spin-polarized current is also known as a consequence of the spin-transfer torque phenomenon and technologically applied in a number of novel spintronic devices such as spin-logic gate [3,4] racetrack memory [5,6], etc. In such devices, the bits of information are stored as magnetic domains in nano-sized magnetic wires and controlled by a spin-polarized current flowed along the wires. Writing of a bit is performed by switching the magnetization direction of a domain by means of a localized external field obtained from for example the Oersted field from a crossing wire [8], or alternatively by means of STT with passing a current from a magnetic nano-element into the wire.

In previous work [9], we proclaimed the possibility to control the STT motion of DWs in perpendicularly magnetized single-layer TbFeCo with very low current density ($\sim 5 \times 10^{10}$ Am$^{-2}$ and high DW speed (30-50 m/s). Thus, the single TbFeCo film with low saturation magnetization, narrow DW width and high anisotropy [10] is very promising for high-performance, low-energy DW devices.

In this paper, we present a demonstration that the writing of domain in TbFeCo can be modulated by adjusting either the writing pulse or external magnetic field that is important for writing process in memory devices.

## 2. Experimental

Nanowire with 300 nm width, 150 μm length was fabricated by means of electron beam lithography and a lift-off technique. A 300 nm thick ZEP 520A e-beam resist was firstly coated on a Si substrate by spin coating. Pattern was then written by a CABL 8000 beam writer system with an accelerated voltage of 30 kV, a beam current of 100 pA. The $Tb_{24}Fe_{68.4}Co_{7.6}$ (20 nm in thickness) film was grown on naturally oxidized Si substrate by RF magnetron sputtering in Ar gas. The base vacuum was $3 \times 10^{-8}$ Torr whereas the Ar pressure was fixed at 2 mTorr. To avoid the ambient oxidization, the film was capped by a 1 nm thick Pt layer. For magnetransport measurement, the wire was modified to be in Hall-bar shape and connected to external electrical sources via a Ti/Al electrodes system made by photolithography and electron beam evaporation (see Fig. 1).

### 3. Results and discussion

First of all, the motion of DWs in the TbFeCo single layer nanowire is demonstrated. As seen from Fig. 1, a pulse current with a frequency of 1.6 MHz (with a pulse width varying from 5 to 40 ns) was flowed along the electrode A-B. This current generated a pulsed Oersted field, which acts to nucleate a tiny domain as the region between the square pad and the wire. Such a tiny domain would be driven to move along the wire by a DC current, $J_{DC}$, passed through the wire in accordance with the spin-transfer torque phenomenon described in the last paragraphs. When each domain passed a Hall cross-bar (either C-D or E-F), it caused a change in the Hall-effect voltage measured between the electrodes. Figure 2(a) illustrates the waveform of the Hall-effect voltage versus time in which the pulse-like of the signal denoted the tiny domains passed the Hall cross-bar. This waveform is similar to the previous results published elsewhere [11,12]. The pulse-like could be qualitatively understood via a schematic drawn in Fig. 2(b): when the domain was moving in the straight wire area, the Hall signal was constant; the Hall signal started increasing at the time $T_1$ as the front-edge wall entered the Hall cross-bar; the Hall-effect voltage remained its high value when the front-edge was drifting inside the Hall bar $T_2 \div T_3$ and switched to low value as the rear-edge wall moved into the cross-bar ($T_3 \div T_4$); the $T_4$ interval denoted the time that the rear-edge wall of the domain passed the Hall bar. Hence, the interval $T_2$-$T_1$ and $T_4$-$T_3$ denoted the velocities of the front-edge and rear-edge walls of the domain inside the cross-bar, respectively whereas the interval $T_3$-$T_1$ reflected the length of the domain [11]. By the same approach, the velocity of the domain traveling in the straight area between the electrodes C-E was determined by measuring the phase delay between the Hall signals at the electrodes C-D and E-F [11]. To obtain such waveform of the Hall-effect signal, minimum DC current values of 0.21 mA was required, corresponding to a critical current density of $J_c = 3.5 \times 10^{10}$ $Am^{-2}$. Such low threshold current density is consistently reduced in comparison with previous result of the TbFeCo nanowire [9] which experimentally proved that the narrower wire was used, less driven current was postulated. This critical current density could be ascribed as one of the lowest value to control magnetic domain walls in the nanowire up to now.

It was found that two intervals $T_2 - T_1$ and $T_4$-$T_3$ were mostly equal each other, certifying that velocities of the front-edge and rear-edge walls were perfectly indistinguishable. Moreover, these velocities were also nicely equivalent to the velocity of the domain running in the straight area, supposing that the motion of the domains through the Hall cross-bar and the straight area of the wire (without Hall bars) was simply identical to each other, pinning effect was relatively weak and the structure of the domains was mostly conservative.

Figure 3 shows the dependence of the front-edge wall's velocity on the driven current density at various external fields. The current dependence of the velocity revealed that the motion of the walls was driven by the spin-transfer torque mechanism and was accordant with the adiabatic model at weak pinning regime in which the velocity is given by [13]:

$$v \sim \sqrt{J_{DC}^2 - J_C^2} \qquad\qquad\qquad (1)$$

Where, $J_{DC}$ is the driven current density, $J_c$ is known as the critical current density. In this model, the current could not drive the wall if the associated spin current was smaller than the critical value [13,14]. It should be noted that the variation of the DW velocity in the Fig. 3 is analogous to previous results on TbFeCo microwires [9] of which the wall velocity was consistent with the law given in the equation (1). The results appeared in Fig. 3 on the other hand indicated that the critical current density was independent on the external field, whilst the wall's velocity was slightly enhanced by the external field. Current-dependence here tended to shift to high velocity regime when the applied field increased that is in agreement with no pinning regime, in which the motion of a DW along a magnetic nanowire was imputed to be distinct characteristics depending on the strength of the applied field and the motion of the wall could be analytically described as $v(H) \sim H$ [14]. It is interesting to note that the effect of the external field on the motion of two walls of the domain became unequal when the field was too high (over 1000 Oe), and the distortion of the domain due to the Oersted field was believed.

The modulation of domain writing in TbFeCo was conducted by adjusting the pulse current - which was used to nucleate the domains in the TbFeCo nanowire. The frequency of the pulse was fixed at 1.6 MHz whereas the pulse width was varied from 5-40 ns. By analyzing the signal of the Hall-effect voltage, the domain length was deducted. Figure 4 presents the pulse-width dependence of the domain size measured at a magnetic field of 50 Oe. It is apparent that the domain size depends on the pulse duration as a linear function and it shifts to higher value as increasing the external field. This can be understood that the pulse current, known as writing current, generated a pulsed Oersted field perpendicular to the surface of the film and the current. The size of the nucleated domain was defined by the time that the Oersted field applied to the film. A longer time of the pulse duration, the Oersted field acted more time on the film and the size of the domain became longer along the wire. Turning back to the waveform of the Hall-effect signal (see Fig. 2(a)), the nucleation of the domains in the wire was synchronic with the periodic pulse (1.6 MHz frequency). These evidences confirmed that the domain stored in the TbFeCo nanowire could be modulated by the writing pulse current. With the minimum pulse duration of 5 ns, it was able to storage a minimum domain size of 630±50 nm.

Figure 5 depicts the variation of the domain length with the external applied field obtained by a writing pulse with pulse duration of 10 ns. A logarithmic variation of the domain length was credibly observed here. It means that the domain writing would be modulated by the external field as:

$$L = L_0 + k \times \log(H+H_0) \hspace{3cm} (2)$$

Here, $L_0$, $H_0$ are initial length and field constants, k is a coefficient, respectively.

This showed that the external field was taken into account to the Oersted field from the writing current to expand the nucleated domains. Extrapolating the logarithmic dependence, the length of the domain would be reduced to 450 nm as the field was zero. This was previously discussed that when an external magnetic field was applied perpendicularly to the film, the nucleated domain - reversed domain, would shrink or expand in order to release the Zeeman energy to compensate the energy dissipation due to the moving DW [15,16]. The domain only stopped shrinking or expanding as the dipolar interaction field was balanced with the external field and the domain was only stable with the length larger than the DW width where the repulsive force due to dipolar field was dominant. The results here suggested that the magnetization in nucleated domain here was in the same direction to the external field.

### 4. Conclusions

Motion of magnetic domain walls in the TbFeCo nanowire by very low current density up to $J_c = 3.5 \times 10^{10}$ Am$^{-2}$ has been demonstrated in terms of magnetotransport measurement. We have established writing of tiny domain (minimum to 630 nm) into the TbFeCo nanowire by combining the nucleation pulse current (30 mA amplitude, 1.6 MHz frequency, 5-40 ns length) and DC driven current. The size of the domain could be modulated by adjusting either external magnetic field or the length of the nucleation pulse. This contribution could be considered as an important note of which the writing process in the DW spin-torque devices must be taken care.

### Acknowledgments

This work was financially supported by the fellowship funded by the Toyota School Foundation.

**Figures caption**

Fig. 1. Scanning electron microscopy (SEM) image of the TbFeCo nanowire device in the electrical circuit for transport measurements.

Figure 2. (a) Waveform of the Hall-effect voltage signal versus time obtained in the external field of 120 Oe, at the critical current $I_{DC} = 0.21$ mA and nucleation pulse width of 10 ns; (b) Interpretation of the pulse-shape Hall-effect signal.

Figure 3. The velocity of the front-edge wall as a function of the driven current density measured at various external fields.

Figure 4. The time modulation of the domain length: domain length is linearly dependent on the duration of the writing pulse.

Figure 5. The field modulation: domain length varies with the external field as a logarithmic function.

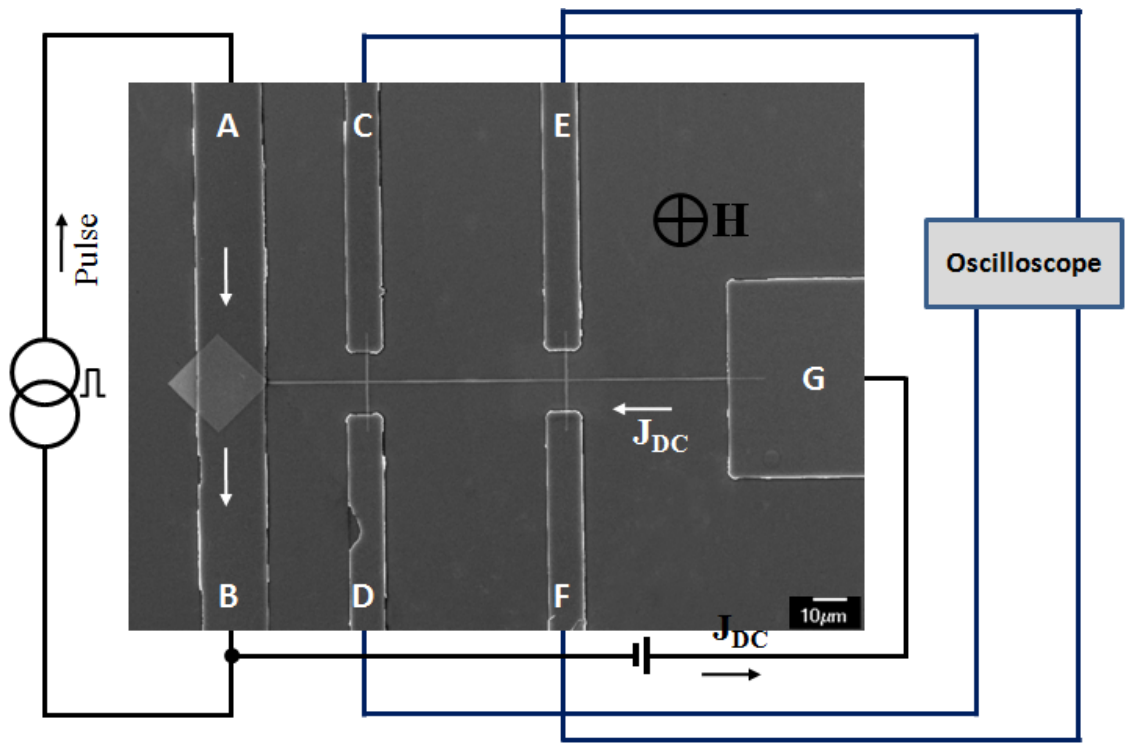

Figure 1.

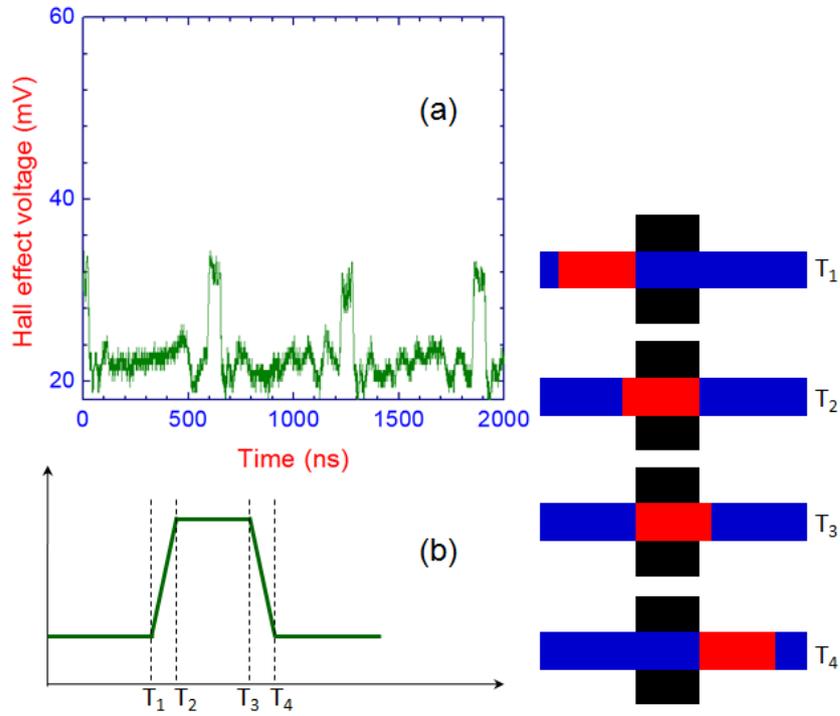

Figure 2.

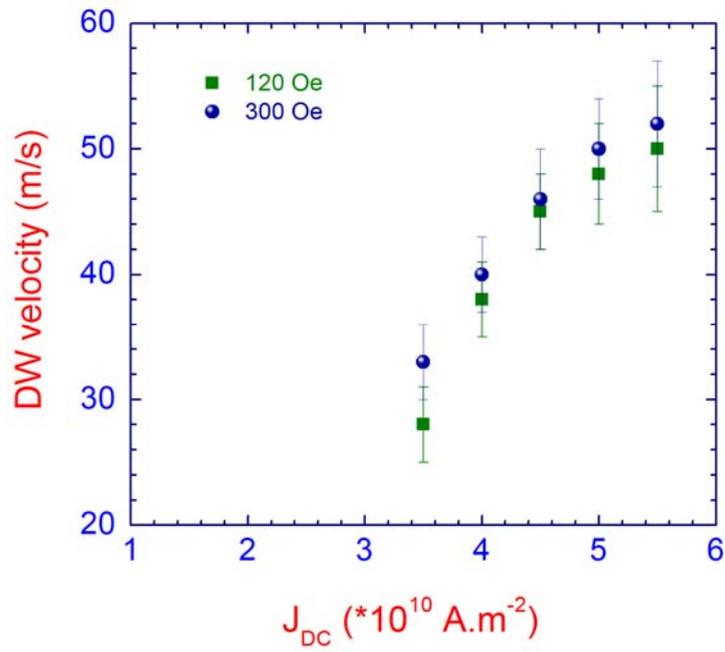

Figure 3

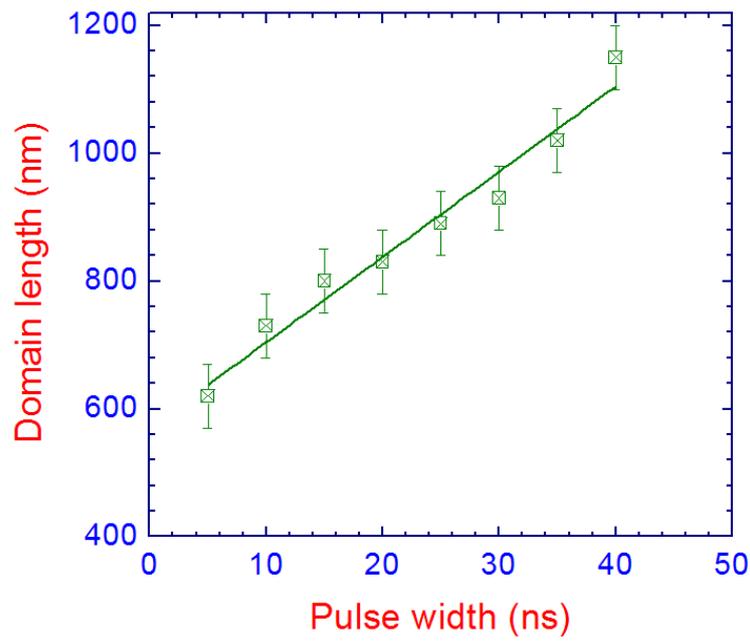

Figure 4

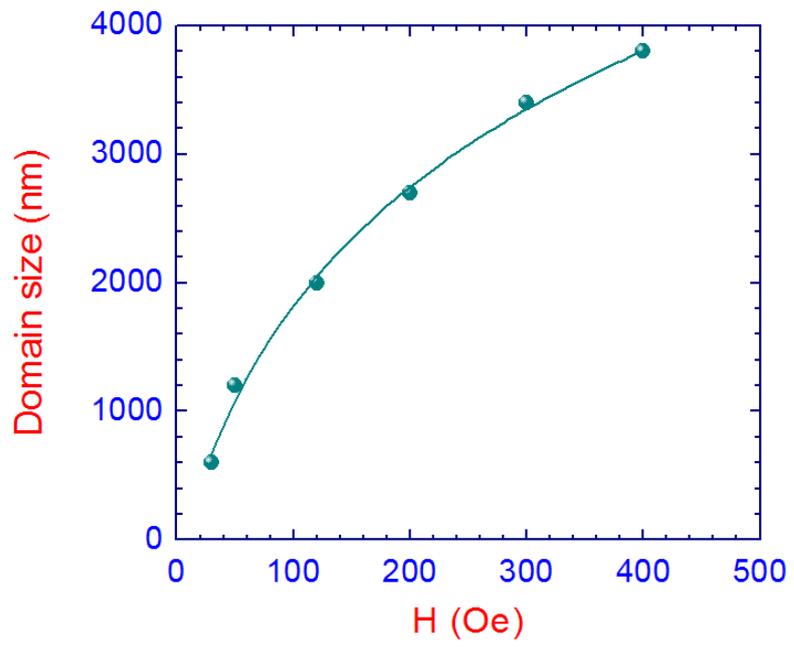

Figure 5.